\RequirePackage[mathlines]{lineno} 
\documentclass[a4paper,11pt]{article}

\usepackage{jheppub} 

\usepackage[T1]{fontenc} 

\usepackage{threeparttable}
\usepackage{multirow}
\usepackage{subfigure}
\usepackage{hyperref}
\usepackage{float}
\let\oldequation\equation
\let\oldendequation\endequation
\renewenvironment{equation}
{\linenomathNonumbers\oldequation}
{\oldendequation\endlinenomath}

\def \jpsi {J/\psi}

\newcommand{\BESIIIorcid}[1]{\href{https://orcid.org/#1}{\hspace*{0.1em}\raisebox{-0.45ex}{\includegraphics[width=1em]{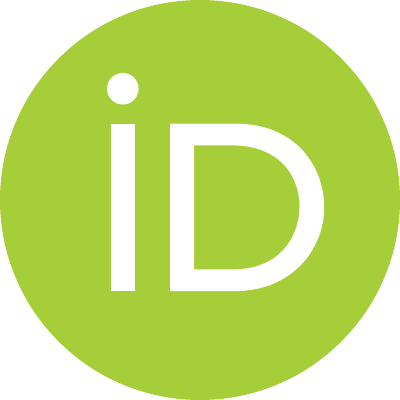}}}}

\begin{document}

\title{Search for the charmonium weak decay $\psi(2S)\to D_s^-\pi^+ + c.c.$ and $\psi(2S)\to D_s^-\rho^+ + c.c.$}

\collaborationImg{\includegraphics[height=30mm,angle=90]{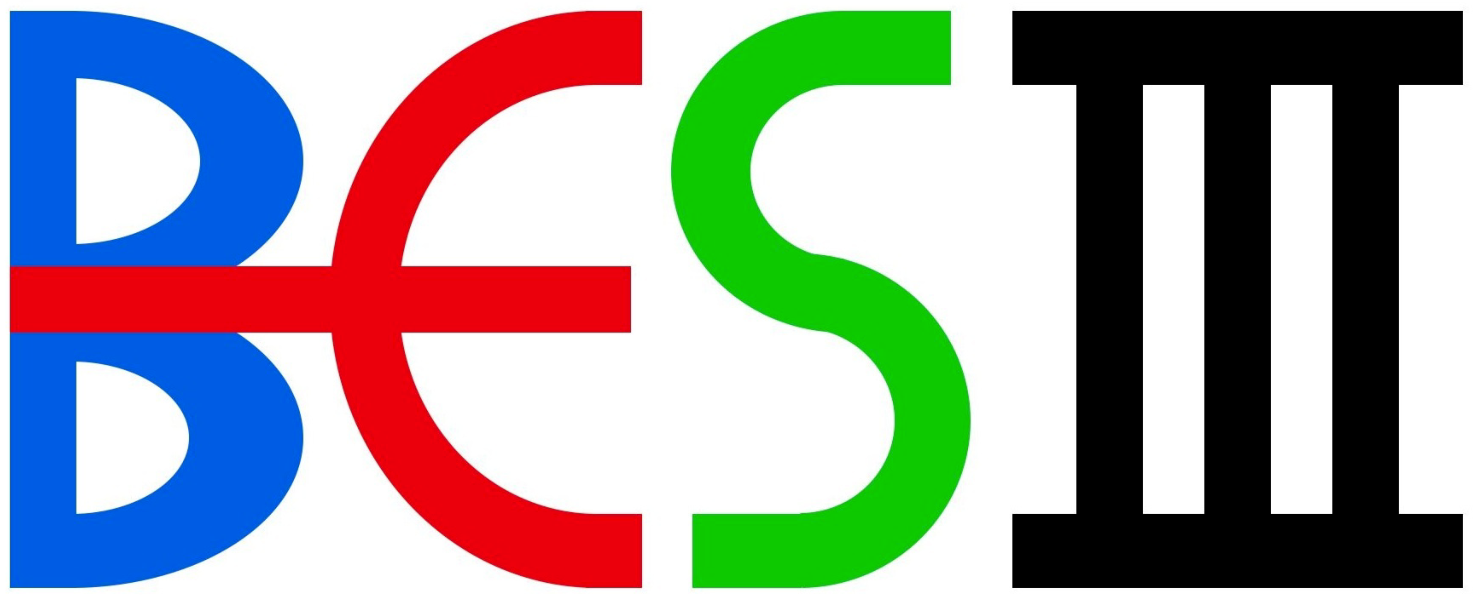}}
\collaboration{The BESIII collaboration}
\emailAdd{besiii-publications@ihep.ac.cn}

\abstract{Using the data sample of $(2712.4\pm14.3)\times 10^6$ $\psi(2S)$ events collected with the BESIII detector at the center-of-mass energy $\sqrt{s}=3.686\ \rm{GeV}$, we search for the weak decays $\psi(2S)\to D_s^-\pi^+ + c.c.$ and $\psi(2S)\to D_s^-\rho^+ + c.c.$ for the first time. Since no signal excess above the background is observed, the upper limits on the branching fractions at the 90\% confidence level are set to be $1.4\times 10^{-6}$ and $7.0\times 10^{-6}$ for $\psi(2S)\to D_s^-\pi^+ + c.c.$ and $\psi(2S)\to D_s^-\rho^+ + c.c.$, respectively.}


\maketitle
\flushbottom


\section{Introduction}
\hspace{1.5em} 

With the $\psi(2S)$ mass being just below the production threshold for a pair of $D$ mesons, the decay of $\psi(2S)$ into $D \bar{D}$ is kinematically forbidden. Decays to a single charmed meson, $D$ or $D_s$, are however still allowed in the Standard Model~(SM). In contrast to the strong decay channels of $\psi(2S)$, its weak decays are suppressed and experimentally challenging, but may offer a unique opportunity to test the SM and search for new physics. As a reference, the SM predicts that the inclusive branching fraction~(BF) for the weak decay of the $J/\psi$ meson into a single charmed meson is at the level of $10^{-8}$ or below~\cite{verma:1990, Sanchis:1994, Sanchis:1993, sharma:1999, wang:2008a, wang:2008b, shen:2008, dhir:2013, ivanov:2015, tian:2017, Sun:2023psi2S_BF_prediction, Meng:2024nyo, Galkin:2026kdf, Wang:2008}; a comprehensive review of the theoretical results and experiment limits for $J/\psi$ charmonium weak decays can be found in Ref.~\cite{Li:2026ced}. These predictions are however beyond the reach of current-generation experiments. 

Nevertheless, several new physics models, such as the Top-color model~\cite{npm:TopColorModel}, the Minimal Supersymmetric SM with or without R-parity~\cite{npm:SuperSymmetryWithRViolation}, and the two-Higgs doublet model~\cite{npm:DoubleHiggs}, predict that the BFs of weak decays of the $J/\psi$ meson can be enhanced by 2-3 orders of magnitude compared to the SM expectations~\cite{Datta:1998yq}, making these decays potentially detectable at the BESIII experiment. While no evidence for new physics has been found so far~---~the BESIII collaboration has already searched for several weak decays of $J/\psi$ into single charmed mesons, including $J/\psi \to D_s^-\pi^+$ and $J/\psi \to D_s^-\rho^+$, and no significant signals have been observed~\cite{bes:DspiandDsrho}. The hadronic weak decays of $\psi(2S)$ remain largely unexplored, the only search up to date is $\psi(2S) \to \Lambda_c^+\bar{\Sigma}^-$~\cite{bes3:LambdaSigma}. For the two charmonium weak decays studied in this work, $\psi(2S)\to D_s^-\pi^+$ and $\psi(2S)\to D_s^-\rho^+$, the SM predicts that their BFs are $1.23^{+0.70}_{-0.72}\times 10^{-10}$ and $1.22^{+0.45}_{-0.32}\times 10^{-9}$~\cite{Sun:2023psi2S_BF_prediction}, respectively. The charge-conjugate modes are always implied throughout this paper.

\begin{figure}
    \centering
    \includegraphics[width=0.5\linewidth]{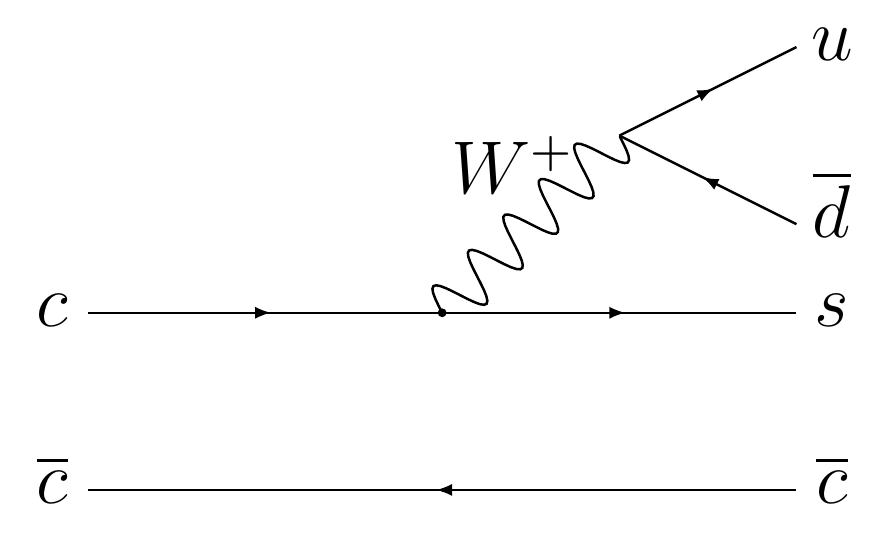}
    \caption{The Feynman diagram of $\psi(2S)\to D_s^-\pi^+$ and $\psi(2S)\to D_s^-\rho^+$ at tree-level.}
    \label{fig:feynman_diag}
\end{figure}

In this study, we search for two weak decays of $\psi(2S)$ utilizing $(2712.4\pm14.3)\times 10^6$ $\psi(2S)$ events collected at $\sqrt{s}=3.686\ \rm{GeV}$ with the BESIII detector~\cite{bes3:psipnum}. As illustrated by the tree-level Feynman diagram shown in Fig.~\ref{fig:feynman_diag}, only one of the charm quarks in the charmonium state decays into a lighter quark through the weak interaction. 
The analysis employs a stepwise closed-box analysis method to avoid potential bias from the analyzers. First, the selection criteria are determined solely based on Monte Carlo (MC) simulated samples. Then, a small data sample, which is randomly selected from the original experimental data sample and contains roughly 10\% of the total events, is used to validate the analysis method and obtain a preliminary result. Finally, the analysis method determined in the previous steps is applied to the entire data sample to obtain the final result.

\section{BESIII detector and data sample}
\label{sec:DetectorAndDataSample}
The BESIII detector~\cite{bes3:2010detector} records symmetric $e^+e^-$ collisions provided by the BEPCII storage ring~\cite{bes3:BEPCII} in the center-of-mass energy range from 1.84 to 4.95~GeV, with a peak luminosity of $1.1 \times 10^{33}\;\text{cm}^{-2}\text{s}^{-1}$ achieved at $\sqrt{s} = 3.773\;\text{GeV}$. BESIII has collected large data samples in this energy region~\cite{bes3:EcmsMea, bes3:Data2}. The cylindrical core of the BESIII detector covers 93\% of the full solid angle and consists of a helium-based multilayer drift chamber~(MDC), a time-of-flight system~(TOF), and a CsI(Tl) electromagnetic calorimeter~(EMC), which are all enclosed in a superconducting solenoidal magnet providing a 1.0~T magnetic field. The solenoid is supported by an octagonal flux-return yoke with resistive plate counter muon identification modules interleaved with steel. 
The charged-particle momentum resolution at $1~{\rm GeV}/c$ is $0.5\%$, and the ${\rm d}E/{\rm d}x$ resolution is $6\%$ for electrons from Bhabha scattering. The EMC measures photon energies with a resolution of $2.5\%$ ($5\%$) at $1$~GeV in the barrel (end-cap) region. The time resolution in the plastic scintillator TOF barrel region is 68~ps, while that in the end-cap region was 110~ps. The end cap TOF system was upgraded in 2015 using multigap resistive plate chamber technology, providing a time resolution of 60~ps, which applies to $\sim 87$\% of the data used in this analysis~\cite{bes3:eTOF, bes3:eTOF2, bes3:eTOF3}.

MC simulated data samples produced with a {\sc geant4}-based~\cite{sof:geant4} software package, which includes the geometric description of the BESIII detector\cite{sof:detvis, sof:detvis2, sof:detvis3} and the detector response, are used to determine selection criteria, detection efficiencies and to estimate backgrounds. 
The simulation models the beam-energy spread and initial-state radiation (ISR) in the $e^+e^-$ annihilations with the generator {\sc kkmc}~\cite{sof:kkmc, sof:kkmc2}. Two types of MC samples are used, the inclusive MC sample and the signal MC sample. The inclusive MC sample contains 2.7 billion $\psi(2S)$ events corresponding to the size of the real data used in the analysis, and it is used to study the backgrounds from $\psi(2S)$ decays. The particle decays within the inclusive MC sample are modeled with {\sc evtgen}~\cite{sof:evtgen, sof:evtgen2} using BFs either taken from the Particle Data Group (PDG)~\cite{pdg:2025}, when available, or otherwise estimated with {\sc lundcharm}~\cite{sof:lundcharm, sof:lundcharm2}. Final-state radiation from charged particles is incorporated using the {\sc photos} package~\cite{sof:photos2}. In contrast, the signal MC samples contain only signal events with specific decay chains. The primary decays $\psi(2S)\to D_s^-\pi^+$ and  $\psi(2S)\to D_s^-\rho^+$ are generated using the VSS and VVS\_PWAVE models, respectively. Here, VSS model describes a vector particle decaying into two scalar particles by incorporating the P-wave nature of the decay, while VVS\_PWAVE model, which describes a vector particle decaying into a vector and a scalar particle, contains several parameters to control the angular distributions of final-state particles. Initially, the parameters are set to produce a pure S-wave decay. The subsequent decays $D_s^-\to\phi e^-\bar{\nu}_e$, $\rho^+\to \pi^0\pi^+$, $\phi\to K^+K^-$ and $\pi^0\to \gamma\gamma$ are simulated with the PHOTOS ISGW2, VSS, VSS and PHSP models, respectively. The ISGW2 model is a constituent quark model for semileptonic $B$ and $D$ meson decays~\cite{Scora:1995ty}. The PHSP model generates decays uniformly over the kinematically allowed phase space. For each decay channel of $\psi(2S)\to D_s^-\pi^+$ and $\psi(2S)\to D_s^-\rho^+$, $2\times 10^5$ signal MC events are generated for each of the three data-taking periods (2009, 2012, and 2021), corresponding to a total of $6\times 10^5$ events per channel.

\section{Event selection and data analysis}
The analysis is performed using the BESIII offline software system~\cite{bes3:boss}.
We reconstruct $D_s^-$ candidates using the semileptonic decay channel $D_s^-\to \phi e^-\bar{\nu}_e$, which has a large BF $(2.34 \pm 0.12)$\%~\cite{pdg:2025}, with $\phi$ reconstructed using $\phi\to K^+K^-$. The product of the branching ratios for this cascade decay is $(1.17\pm 0.06)$\%. This decay channel is used in preference to a hadronic decay of the $D_s^-$ meson since the $\psi(2S)$ meson predominantly decays to hadrons through the strong interaction, and a semileptonic final state should therefore have a lower background rate.
The $\rho^+$ meson candidates are reconstructed via $\rho^+\to \pi^+\pi^0$ with $\pi^0\to \gamma\gamma$. 
The reconstructed decay chains of $\psi(2S)\to D_s^-\pi^+$ and $\psi(2S)\to D_s^-\rho^+$ contain four charged final-state particles, corresponding to four charged tracks in the detector. For signal events, these four charged tracks must be detected in the MDC within $|\cos\theta|<0.93$, where $\theta$ is the polar angle with respect to the MDC symmetry axis. The distance of closest approach of each charged track to the interaction point should be smaller than 10 cm along the beam direction and 1 cm in the plane perpendicular to the beam. Furthermore, since the $\psi(2S)$ meson is electrically neutral, the net charge of the selected charged tracks must be zero.

Particle identification (PID) for charged tracks combines the measurements of specific ionization energy loss~(d$E$/d$x$) in the MDC and the time-of-flight measured by TOF to obtain the likelihoods for each particle hypothesis, $L(h)$ with $h=K, \pi, e$. For charged pion candidates we require $L(\pi)>0, L(\pi)>L(K)$ and $L(\pi)>L(e)$. For the kaon candidates we require $L(K)>0$ and $L(K)>L(\pi)$. For electron candidates, the requirements are $L(e)>0.001$ and $L(e)/(L(e) + L(\pi) + L(K))>0.8$. To further suppress $e/\pi$ misidentification, we additionally require $0.86<E/p<1.03$ for $e$ candidates, where $E$ is the energy deposit in the EMC and $p$ is the momentum measured by the MDC; the criteria has been optimized according to the Punzi significance~\cite{Punzi:2003bu}. Both $\psi(2S)\to D_s^-\pi^+$ and $\psi(2S)\to D_s^-\rho^+$ decays produce the same charged final-state particles, $K^+, K^-, \pi^+, e^-$; therefore, two $K$ candidates, one $\pi$ candidate and one $e$ candidate are required for each signal event.

The photon candidates are identified with energy deposited in the EMC. The energy of a good photon candidate must exceed 25~MeV in the EMC barrel region~($|\cos\theta|<0.8$) or 50~MeV in the EMC end-cap region~($0.86<|\cos\theta|<0.92$). To suppress noise from background showers, photon candidates must be detected within 700~ns of the event collision. 
To suppress the shower backgrounds due to hadronic interactions or bremsstrahlung of charged tracks in the EMC, each photon candidate is further required to have an opening angle larger than $10^{\circ}$ from the nearest charged track.
For the decay $\psi(2S)\to D_s^-\rho^+$, two good photon candidates are required.

After applying PID and photon selection criteria, we reconstruct the decay chains $\psi(2S)\to D_s^-\pi^+$ and $\psi(2S)\to D_s^-\rho^+$, where the $D_s^-$ subsequently decays into $\phi e^-\bar{\nu}_e$, with $\phi\to K^+K^-$. The $\phi$ candidates are selected by requiring the invariant mass of the $K^+K^-$ pair to be within the range $(1.005, 1.035)\,\mathrm{GeV}/c^2$, which corresponds to $\pm 3\sigma$ around the nominal $\phi$ mass.

In the decay $\psi(2S)\to D_s^-\rho^+$, the $\rho^+$ meson is reconstructed via $\rho^+\to \pi^+\pi^0$. The $\pi^0$ candidates are reconstructed from photon pairs. A one-constraint kinematic fit is performed to constrain the $\gamma\gamma$ invariant mass to the nominal $\pi^0$ mass. For events with multiple $\pi^0$ candidates, the one with the smallest $\chi^2$ is selected, provided that $\chi^2<200$. Subsequently, the $\rho^+$ candidates are reconstructed by combining the selected $\pi^0$ meson with a $\pi^+$ track. The $\pi^+\pi^0$ invariant mass is required to be within the range $(0.61, 0.93)\,\mathrm{GeV}/c^2$, which corresponds to approximately $\pm 1\sigma$ around the nominal $\rho$ mass.

\begin{figure*}[htbp]\centering
	\subfigure[]{\includegraphics[width=0.49\textwidth]{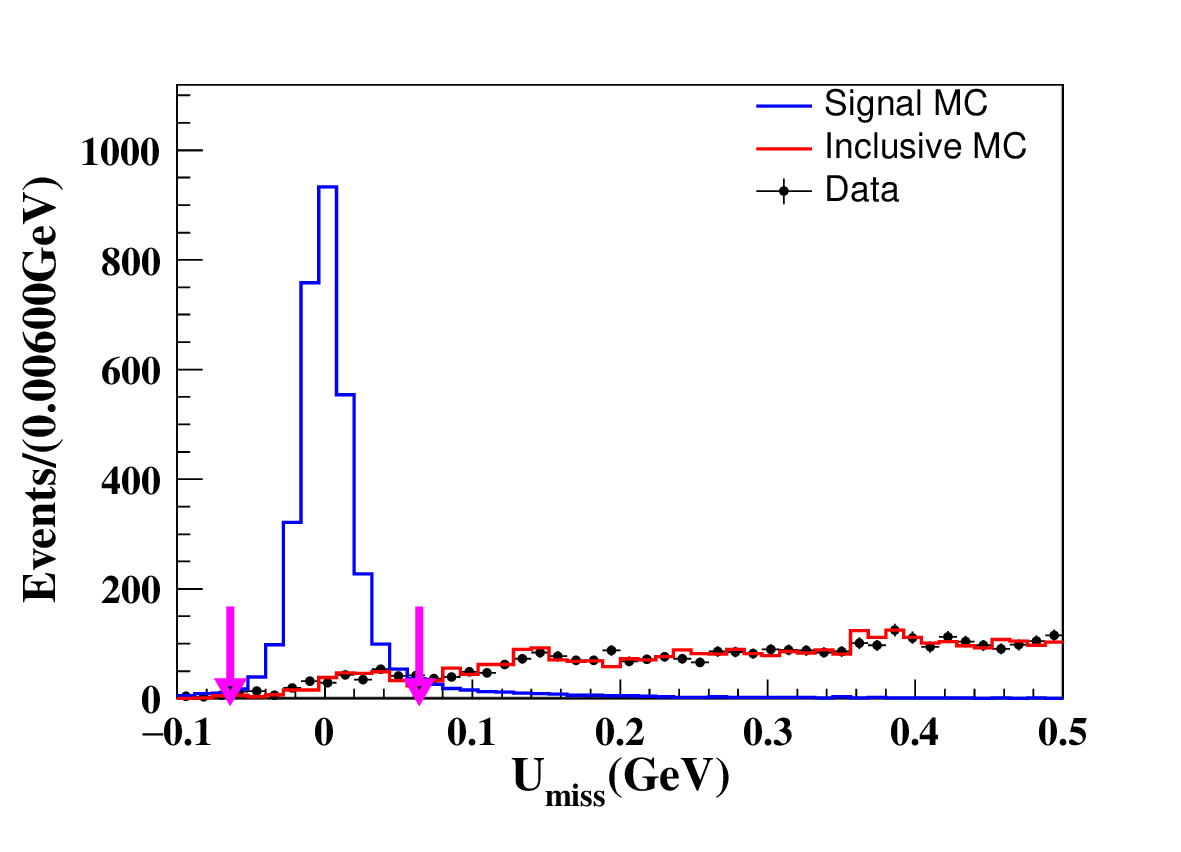}}
    \subfigure[]{\includegraphics[width=0.49\textwidth]{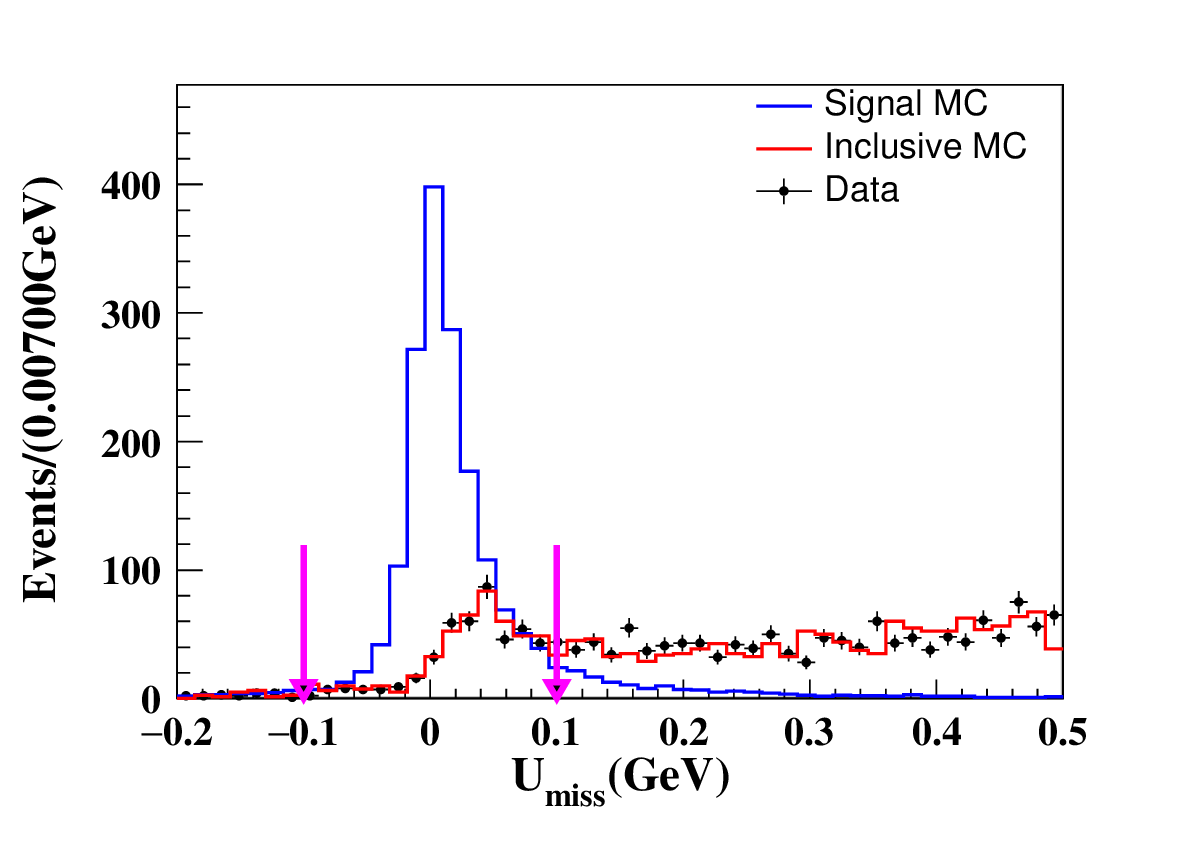}}
	\caption{The $U_{\mathrm{miss}}$ distribution for the decay channels (a)~$\psi(2S)\to D_s^-\pi^+$ and (b)~$\psi(2S)\to D_s^-\rho^+$. The black dots with error bars are data. The signal and inclusive MC distributions, shown as the blue and red lines respectively, are scaled to have the same integral area as the data for visual comparison. The magenta arrows indicate the defined signal regions.}
	\label{fig: pmiss_and_umiss}
\end{figure*}

To suppress backgrounds that do not contain neutrinos, we require $|\vec{p}_{\mathrm{miss}}|>0.02$~GeV/c, optimized according to the Punzi significance~\cite{Punzi:2003bu}, and $|U_{\mathrm{miss}}|<0.064~\mathrm{GeV}$ for $\psi(2S)\to D_s^-\pi^+$ mode, $|U_{\mathrm{miss}}|<0.10~\mathrm{GeV}$ for $\psi(2S)\to D_s^-\rho^+$ mode. The variables $\vec{p}_{\mathrm{miss}}$ and $U_{\mathrm{miss}}$ are defined as:
\begin{eqnarray}
	\vec{p}_{\mathrm{miss}} = \vec{p}_{\psi(2S)} - \vec{p}_{K^+} - \vec{p}_{K^-} - \vec{p}_{e^-} - \vec{p}_{\pi^+(\rho^+)},
\end{eqnarray}
\begin{eqnarray}
	E_{\mathrm{miss}} = E_{\psi(2S)} - E_{K^+} - E_{K^-} - E_{e^-} - E_{\pi^+(\rho^+)},
\end{eqnarray}
\begin{eqnarray}
	U_{\mathrm{miss}} = E_{\mathrm{miss}} - |\vec{p}_{\mathrm{miss}}|c,
\end{eqnarray}
where $E_i$ and $\vec{p}_i$ denote the energy and momentum of particle $i$, with $i\in\{\psi(2S)$, $K^+$, $K^-$, $e^-$, $\pi^+(\rho^+)\}$. Since the $\psi(2S)$ is produced at rest in the $e^+e^-$ center-of-mass frame at $\sqrt{s} = 3.686$~GeV, the initial state information $\vec{p}_{\psi(2S)} = 0$ and $E_{\psi(2S)} = \sqrt{s}$ is used. The requirement on $|\vec{p}_{\mathrm{miss}}|$ is used to suppress backgrounds containing only detectable particles, and the requirement on $U_{\mathrm{miss}}$ is used to identify the neutrino candidates. As the neutrino mass is negligible, the value of $U_{\mathrm{miss}}$ is expected to be zero for signal events. The $U_{\mathrm{miss}}$ distributions are shown in Fig.~\ref{fig: pmiss_and_umiss}.

Following the preceding selections, an additional selection criterion is used to further suppress the background in the $\psi(2S) \to D_s^- \rho^+$ channel. Studies of the inclusive MC sample indicate that this channel suffers from higher background contamination than $\psi(2S) \to D_s^- \pi^+$. The dominant contribution arises from final states with multiple $\pi^0/\gamma$ such as $K^+K^-\pi^+\pi^-\pi^0 \pi^0$ and $K^+K^-\pi^+\pi^-\pi^0 \gamma$. These background events enter the signal region primarily through $e/\pi$ misidentification. To suppress this background, we define $E_{\gamma}^ {\mathrm{rest}}$ as the total energy of all photons except the two photons from the $\pi^0$ meson in the $\rho^+$ decay, and require $E_{\gamma}^ {\mathrm{rest}}<0.12$~GeV. The criteria has been optimized according to the Punzi significance~\cite{Punzi:2003bu}.

\begin{figure*}[htbp]\centering
	\subfigure[]{\includegraphics[width=0.49\textwidth]{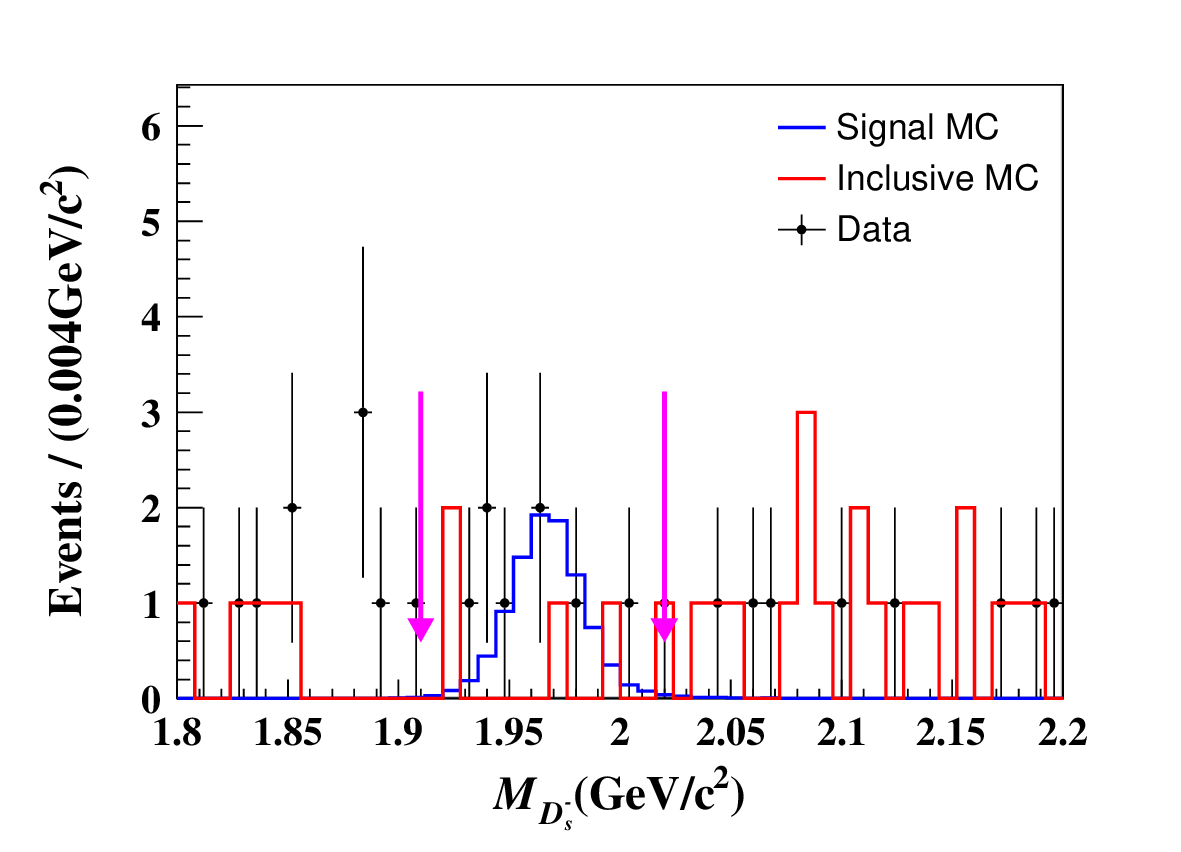}}
	\subfigure[]{\includegraphics[width=0.49\textwidth]{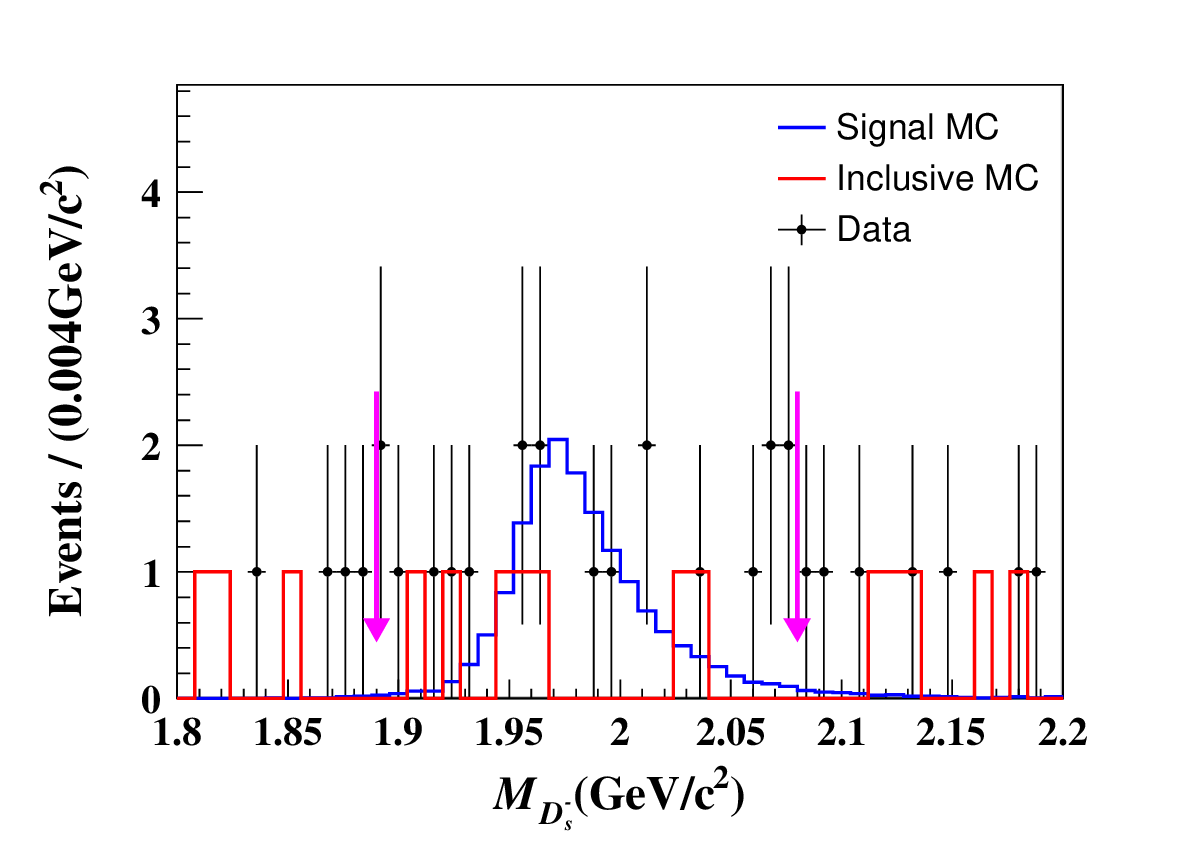}}
	\caption{The invariant mass distributions $M_{D_s^-}$ for the decay channels (a)~$\psi(2S)\to D_s^-\pi^+$ and (b)~$\psi(2S)\to D_s^-\rho^+$. The black dots with error bars are data; while the blue and red lines are the signal and inclusive MC samples, respectively. The signal MC distributions have been scaled to the calculated upper limits of BFs. The magenta arrows indicate the defined signal regions.}
	\label{fig: M_Ds}
\end{figure*}

Finally, we identify the $D_s^-$ candidates to extract the signal events. Because the decay chain of $D_s^-$ includes a neutrino, the $D_s^-$ cannot be fully reconstructed via $\phi e^- \bar{\nu}_e$. Instead, we identify the $D_s^-$ candidate using the recoil method. The energy ($E_{D_s^-}$) and momentum ($\vec{p}_{D_s^-}$) of the $D_s^-$ candidate are calculated by
\begin{eqnarray}
	E_{D_s^-} = E_{\psi(2S)} - E_{\pi^+/\rho^+},
\end{eqnarray}
\begin{eqnarray}
	\vec{p}_{D_s^-} = \vec{p}_{\psi(2S)} - \vec{p}_{\pi^+/\rho^+}.
\end{eqnarray}
The invariant mass, $M_{D_s^-}$, is then obtained as $M_{D_s^-} = \sqrt{E_{D_s^-}^2/c^4-|\vec{p}_{D_s^-}|^2/c^2}$. For signal extraction, the $M_{D_s^-}$ window is set to be the $\pm 3\sigma$ ranges around the $D_s$ nominal mass, which are exactly (1.91, 2.02)~GeV/$c^2$ and (1.89, 2.08)~GeV/$c^2$ for $\psi(2S) \to D_s^- \pi^+$ and $\psi(2S) \to D_s^- \rho^+$ modes, respectively. The $M_{D_s^-}$ distributions for data and MC samples are shown in Fig.~\ref{fig: M_Ds}. The detection efficiencies after event selection are determined using the signal MC samples. They are 21.7\% for the $\psi(2S)\to D_s^-\pi^+$ decay and 7.1\% for the $\psi(2S)\to D_s^-\rho^+$ decay.

\section{Background study}
After applying the above selection criteria, most background events in the data sample have been removed. However, a study is still necessary to determine the residual background. The background events are categorized into two classes: the backgrounds from $\psi(2S)$ decays, and the continuum backgrounds. The first category is analyzed using the inclusive $\psi(2S)$ MC sample, while the second is estimated with data collected at $\sqrt{s}=3.650,\ 3.682,$ and $3.773$~GeV.

The number of background events from $\psi(2S)$ decays in the signal region is estimated as
\begin{eqnarray}
	N_{\mathrm{bkg1}} = N^{\mathrm{obs}}_{\mathrm{bkg1}}\times f_1,\ f_1=\frac{N^{\mathrm{data}}_{\psi(2S)}}{N^{\mathrm{MC}}_{\psi(2S)}},
\end{eqnarray}
where $N_{\mathrm{bkg1}}$ is the estimated background yield, $N^{\mathrm{obs}}_{\mathrm{bkg1}}$ is the number of surviving events in the inclusive MC sample within the $M_{D_s}$ signal window, $f_1$ is the scaling factor, defined as the ratio of the number of $\psi(2S)$ events in the data to that in the inclusive MC sample. The numbers of background events from $\psi(2S)$ decays are estimated to be $4.9\pm2.2$ and $6.9\pm2.6$ for $\psi(2S) \to D_s^- \pi^+$ and $\psi(2S) \to D_s^- \rho^+$ channels,  respectively.

The continuum background contribution is estimated by taking the average of the scaled backgrounds from data samples at various $\sqrt{s}$ energy points and  scaled by a factor $f_2$ that accounts for the $\sqrt{s}$ and luminosity dependence, as shown in Eq.~\ref{eq:bkg2}.
\begin{eqnarray}
\label{eq:bkg2}
	N_{\mathrm{bkg2}} = \frac{\sum_i N^{\mathrm{obs},i}_{\mathrm{bkg2}}}{\sum_i 1/f_2^i}, \ f^i_2=\frac{L_{\psi(2S)}}{L_{i}}\frac{s_i}{s_{\psi(2S)}},
\end{eqnarray}
where $N^{\mathrm{obs},i}_{\mathrm{bkg2}}$ denotes the number of surviving events within the $M_{D_s}$ signal window in the continuum data sample at $\sqrt{s_i}$, with $\sqrt{s_i}=3.650, 3.682$ or $3.773$~GeV, $L_{i}$ and $L_{\psi(2S)}$ are the corresponding integrated luminosities, and $s_i$ and $s_{\psi(2S)}$ are the squared center-of-mass energies. With Eq. \ref{eq:bkg2}, the numbers of continuum background events are estimated to be $0.7\pm0.7$ and $5.5\pm1.9$ for $\psi(2S) \to D_s^- \pi^+$ and $\psi(2S) \to D_s^- \rho^+$ modes, respectively. The total number of background events, obtained by summing the two background contributions, is estimated to be $5.6\pm2.3$ for $\psi(2S) \to D_s^- \pi^+$ decays and $12.4\pm3.2$ for $\psi(2S) \to D_s^- \rho^+$ decays.

\section{Systematic uncertainties}
\label{sec:SysUnc}
The systematic uncertainties in the measurements of the BFs of $\psi(2S) \to D_s^- \pi^+$ and $\psi(2S) \to D_s^- \rho^+$ decays are mainly from the following sources: the MC generator model, tracking and PID efficiencies for charged particles, intermediate BFs, the number of $\psi(2S)$ events, MC statistics, photon detection efficiencies, and the selection criteria. This section describes the evaluation methodology for each source of systematic uncertainty, and the corresponding numerical values are summarized in Table~\ref{tab:systematic_uncertainties}.

\begin{itemize}
	\item \emph{MC generator model}. To estimate the systematic uncertainty from MC generator, alternative models or modified parameters are used to generate new signal MC samples, from which the corresponding detection efficiencies are evaluated. For $\psi(2S)\to D_s^{-}\pi^{+}$, as described in Section~\ref{sec:DetectorAndDataSample}, the decay is simulated using the VSS model. Since this is a purely P-wave decay and the momentum distribution is well-determined by energy-momentum conservation, the systematic uncertainty introduced by the VSS model is expected to be negligible. For $\psi(2S)\to D_s^-\rho^+$, the nominal MC generator model is VVS\_PWAVE. To account for potential P-wave and D-wave contributions, we vary the parameters to generate P-wave and D-wave decay samples and use them as alternative signal MC sets. The standard deviation of the detection efficiencies of these MC samples is taken as the systematic uncertainty, which is measured to be 8.6\%.
	
	\item \emph{Tracking and PID}. The systematic uncertainties from tracking and PID for kaons and pions are estimated using the control sample of $\psi(3770)\to D^0\bar{D}^0(D^+D^-)$. The hadronic decay modes $D^0\to K^-\pi^+, K^-\pi^+\pi^+\pi^-$ (and its charge conjugate $\bar{D}^0\to K^+\pi^-, K^+\pi^-\pi^-\pi^+$) as well as $D^+\to K^-\pi^+\pi^+$ versus $D^-\to K^+ \pi^-\pi^-$ are used as control channels. In these samples, a $K$ or $\pi$ meson is intentionally omitted from the reconstruction to mimic tracking inefficiencies and evaluate the associated systematic uncertainties, and the systematic uncertainty is derived from the difference in tracking efficiencies between data and MC~\cite{bes3:tracking_PID_uncertainty}. 
    
    In addition, the systematic uncertainty from $e$ tracking is studied using the control sample of radiative Bhabha processes $e^+e^-\to e^+e^-\gamma$, while the systematic uncertainty of $e$ PID is studied using a mixed control sample of radiative Bhabha events at $J/\psi$ energy point and $J/\psi\to e^+e^-\gamma$ decays~\cite{bes3:e_tracking_PID_uncertainty}. According to these studies, the systematic uncertainties from tracking and PID are both assigned to be 1\% per charged track.
	
	\item \emph{$\gamma$ detection}. The systematic uncertainty due to $\gamma$ detection is estimated to be 1\% per photon, using the control sample $J/\psi\to \rho^0\pi^0$ and $e^+e^-\to \gamma\gamma$~\cite{bes3:gamma_detection_uncertainty}.
	
	\item \emph{Intermediate BF}. The intermediate BF is defined as $B_{\mathrm{inter}}=B(D_s^-\to \phi e^-\bar{\nu}_e)\cdot B(\phi\to K^+K^-)$ for the $\psi(2S) \to D_s^- \pi^+$, and $B_{\mathrm{inter}}=B(D_s^-\to \phi e^-\bar{\nu}_e)\cdot B(\phi\to K^+K^-)\cdot B(\rho^+\to \pi^+\pi^0)\cdot B(\pi^0\to \gamma\gamma)$ for the $\psi(2S) \to D_s^- \rho^+$. These values are used to calculate the BFs of the two decay channels. The uncertainties of the intermediate BFs for both channels are assigned to be 5.3\%~\cite{pdg:2025}.
	
	\item \emph{Number of $\psi(2S)$ events}. The total number of $\psi(2S)$ events collected with the BESIII detector is $(2712.4\pm14.3)\times 10^6$. Following Ref.~\cite{bes3:psipnum}, a relative uncertainty of 0.5\% is assigned to account for this source.
	
	\item \emph{MC statistics}. The systematic uncertainties due to the limited statistics of the signal MC samples are estimated to be 0.5\% and 0.9\% for $\psi(2S)\to D_s^-\pi^+$ and $\psi(2S)\to D_s^-\rho^+$, respectively.
	
	\item \emph{$U_{\mathrm{miss}}, |\vec{p}_{\mathrm{miss}}|, E/P$ and $E_{\gamma}^ {\mathrm{rest}}$ requirements}. 
The systematic uncertainties associated with these selection requirements are evaluated using the control sample $\psi(3770)\to D^0\bar{D}^0$, $D^0\to K^-e^+\nu_e$ and $\bar{D}^0\to K^+\pi^-(\pi^0)$. For each variable, we select a clean control sample from data and calculate its selection efficiency, and then compare it with the corresponding efficiency obtained from the MC sample. The relative difference between the efficiencies is assigned as the systematic uncertainty.

Based on this method, the systematic uncertainties from the $|\vec{p}_{miss}|, E/P$ and $E_{\gamma}^ {\mathrm{rest}}$ requirements are estimated to be 0.1\%, 1.3\% and 4.5\%, respectively. The uncertainties from the $U_{\mathrm{miss}}$ requirement are determined to be 4.3\% and 3.3\% for $\psi(2S)\to D_s^-\pi^+$ and $\psi(2S)\to D_s^-\rho^+$, respectively. 

	\item \emph{$M_{K^+K^-}$ requirement}. The systematic uncertainty from $M_{K^+K^-}$ requirement is studied using a control sample of $J/\psi\to \phi\eta$ decays. The selection efficiency of control sample is determined by fitting the $M_{K^+K^-}$ distribution in the control sample. The relative difference of efficiencies between the data and MC sample is taken as the systematic uncertainty. This uncertainty is determined to be 0.5\%.
	
	\item \emph{$M_{\pi^+\pi^0}$ requirement}. Similarly, the systematic uncertainty from $M_{\pi^+\pi^0}$ requirement is estimated using a control sample of $J/\psi\to\rho^+\pi^-$ decays. By extracting the signal yields via fitting and comparing the selection efficiencies between the data and MC sample, the systematic uncertainty is determined to be 2.4\%.
	
	\item \emph{$M_{D_s^-}$ requirement}. The systematic uncertainty from $M_{D_s^-}$ requirements are evaluated by varying the $M_{D_s^-}$ signal regions by $\pm 1\sigma$, which corresponds to $\pm 0.018$ GeV/$c^2$ for $\psi(2S)\to D_s^-\pi^+$ and $\pm 0.032$ GeV/$c^2$ for $\psi(2S)\to D_s^-\rho^+$. The relative variations of detection efficiencies are taken as the systematic uncertainties. They are assigned as 1.4\% and 2.2\% for $\psi(2S)\to D_s^-\pi^+$ and $\psi(2S)\to D_s^-\rho^+$ decays, respectively.
\end{itemize}

Finally, assuming that all sources of systematic uncertainty are uncorrelated, the total systematic uncertainty is obtained by taking the square root of the sum of the individual contributions squared. The total systematic uncertainties are 9.2\% for $\psi(2S)\to D_s^-\pi^+$ and 13.6\% for $\psi(2S)\to D_s^-\rho^+$.

\begin{table}
	\caption{The systematic uncertainties~(in \%) from different sources. The mark "-" means this source has no effect on the specific decay channel.}
	\label{tab:systematic_uncertainties}
	\setlength{\abovecaptionskip}{1.2cm}
	\setlength{\belowcaptionskip}{0.2cm}
	\begin{center} 
		\footnotesize
		\vspace{-0.0cm}
		\begin{tabular}{l|cc}
			\hline \hline
			Source & $\psi(2S)\to D_s^-\pi^+$ & $\psi(2S)\to D_s^-\rho^+$\\
			\hline
			MC generator model & - & 8.6\\
			Tracking & 4.0 & 4.0\\
			PID & 4.0 & 4.0\\
			$\gamma$ detection & - & 2.0\\
            $B_{\mathrm{inter}}$ & 5.3 & 5.3 \\
            $N_{\psi(2S)}$ & 0.5 & 0.5\\
			MC statistics & 0.5 & 0.9\\
			$E/P$ & 1.3 & 1.3\\
            $p_{\mathrm{miss}}$ & 0.1 & -\\
			$U_{\mathrm{miss}}$ & 4.3 & 3.3\\
            $E_{\gamma}^ {\mathrm{rest}}$ & - & 4.5\\
			$M_{K^+K^-}$ & 0.5 & 0.5\\
			$M_{\pi^+\pi^0}$ & - & 2.4\\
			$M_{D_s^-}$ & 1.4 & 2.2\\
			\hline
			total & 9.2 & 13.6\\
			\hline \hline
		\end{tabular}
	\end{center}
\end{table}

\section{Results}
After applying the previous selection criteria to the full $\psi(2S)$ data sample, 9 events are observed in the signal region for $\psi(2S)\to D_s^-\pi^+$ and 19 events for $\psi(2S)\to D_s^-\rho^+$. For a statistically significant signal, the BF of a decay mode would be determined using
\begin{eqnarray}
    \label{eq_BF}
    B = \frac{N_{\mathrm{sig}}}{N_{\psi(2S)}\times\epsilon\times B_{\mathrm{inter}}}.
\end{eqnarray}

\begin{figure}\centering
	\subfigure[]{\includegraphics[width=0.49\textwidth]{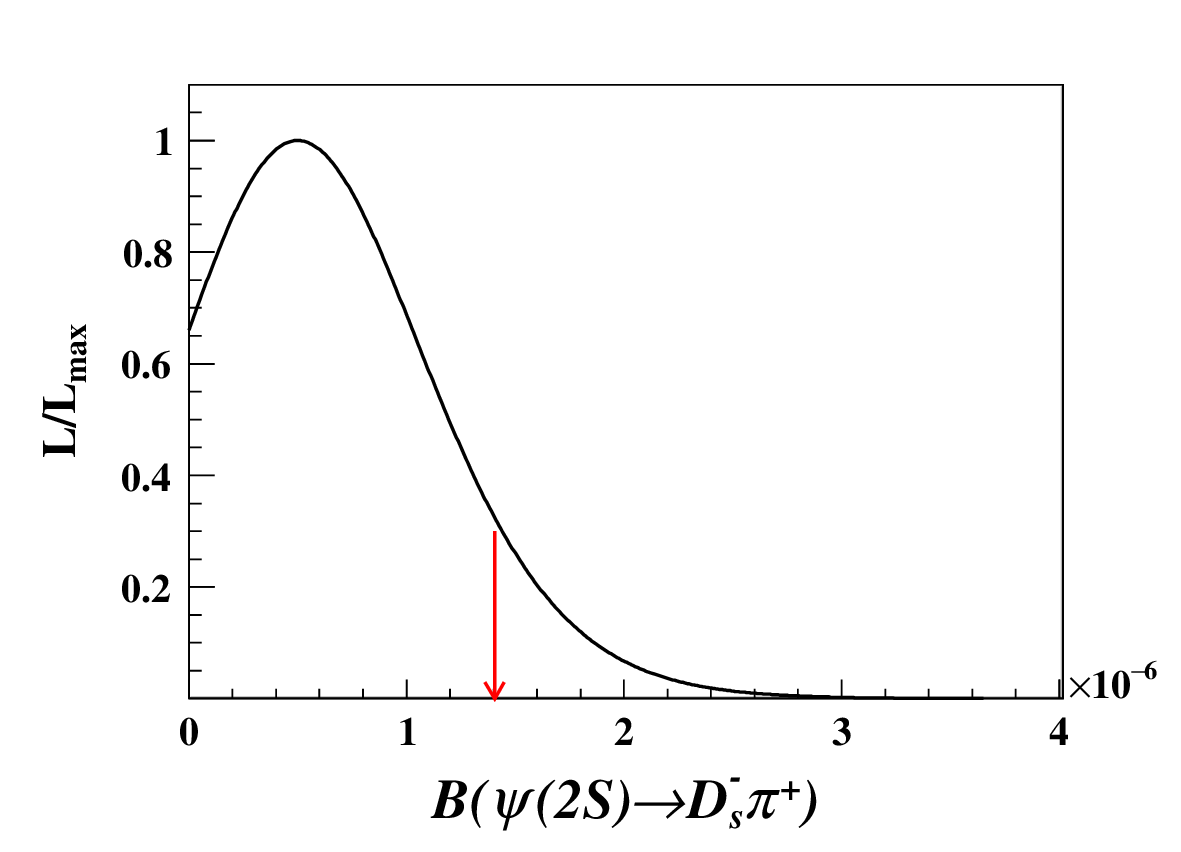}}
	\subfigure[]{\includegraphics[width=0.49\textwidth]{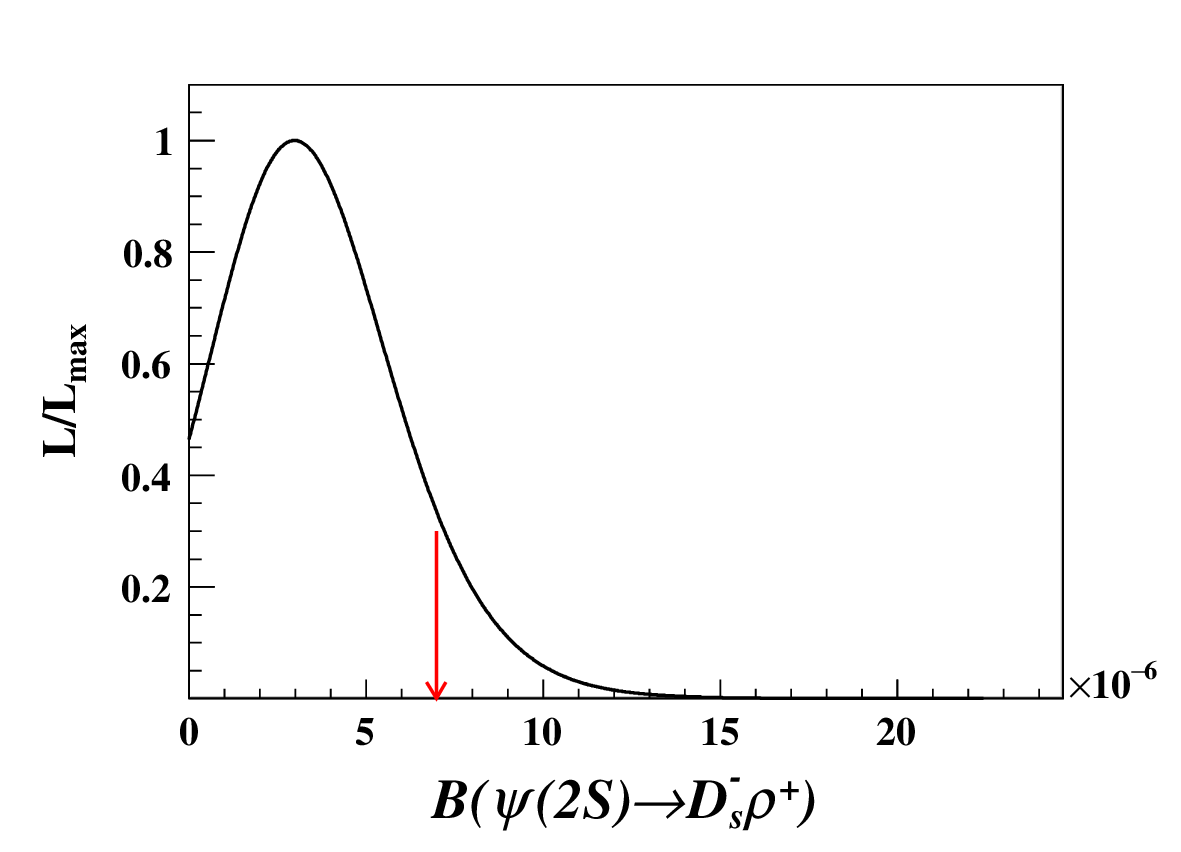}}
	\caption{The normalized likelihoods for (a) $\psi(2S)\to D_s^-\pi^+$ and (b) $\psi(2S)\to D_s^-\rho^+$. The red arrows mark the upper limits of BFs at the 90\% C.L..}
	\label{fig:likelihood_Lb}
\end{figure}

Where $N_{\mathrm{sig}}$ is the number of signal events, $N_{\psi(2S)}$ is the total number of $\psi(2S)$ events, $\epsilon$ is the detection efficiency, and $B_{\mathrm{inter}}$ is the product of intermediate branching fractions. However, because no statistically significant excess above the background is observed, Equation~\ref{eq_BF} cannot be applied directly. 
Instead, upper limits on the BFs of $\psi(2S)\to D_s^-\pi^+$ and $\psi(2S)\to D_s^-\rho^+$ are determined using likelihood functions constructed following the Profile Likelihood method~\cite{profileLikelihood}.
The likelihood functions are defined as
\begin{eqnarray}
	&L&(B,\epsilon_{\mathrm{sig}},N_{\psi(2S)},N_{\mathrm{bkg1}},N_{\mathrm{bkg2}})=\\
	&P&(N_{\mathrm{obs}}|N_{\psi(2S)}\cdot B\cdot\epsilon_{\mathrm{sig}}\cdot B_{\mathrm{inter}}+N_{\mathrm{bkg1}}+N_{\mathrm{bkg2}}) \nonumber \\
	&\cdot& G(\epsilon_{\mathrm{sig}}|\epsilon_{\mathrm{sig}}^{\mathrm{MC}}, \epsilon_{\mathrm{sig}}^{\mathrm{MC}}\cdot\sigma_{\mathrm{sys}}) \nonumber \\
	&\cdot& P(N^{\mathrm{obs}}_{\mathrm{bkg1}}|N_{\mathrm{bkg1}}/f_1)\cdot\prod_iP(N^{\mathrm{obs},i}_{\mathrm{bkg2}}|N_{\mathrm{bkg2}}/f^i_2), \nonumber
\end{eqnarray}
where $N_{\mathrm{obs}}$ is the number of observed events in the signal region, $B$ is the signal BF, $\epsilon_{\mathrm{sig}}$ is the detection efficiency, and $B_{\mathrm{inter}}$ is the intermediate BF. $N_{\mathrm{bkg1}}$ and $N_{\mathrm{bkg2}}$ are the expected numbers of background events from $\psi(2S)$ decay and continuum production, respectively. Here we assume the $\epsilon_{\mathrm{sig}}$ obeys Gaussian distribution whose mean and deviation are the MC-determined efficiency $\epsilon_{\mathrm{sig}}^{\mathrm{MC}}$ and its systematic uncertainty $\epsilon_{\mathrm{sig}}^{\mathrm{MC}}\cdot \sigma_{\mathrm{sys}}$, where $\sigma_{\mathrm{sys}}$ is the total relative systematic uncertainty of the detection efficiency evaluated in Section~\ref{sec:SysUnc}, i.e.\ 9.2\% for $\psi(2S)\to D_s^-\pi^+$ and 13.6\% for $\psi(2S)\to D_s^-\rho^+$, while $N_{\mathrm{obs}}$, $N_{\mathrm{bkg1}}$ and $N_{\mathrm{bkg2}}$ follow the Poisson distribution. Using the Profile likelihood method, we obtain the likelihood distribution $L(B)$, and the upper limit of BF is determined via scanning $L(B)$ in steps of $10^{-8}$ and integrating the distribution until the accumulated likelihood reaches 90\% of the total. The likelihood distribution $L(B)$ and the scanning result are both shown in Fig.~\ref{fig:likelihood_Lb}. At the 90\% C.L., the resulting upper limits are determined to be $1.4\times10^{-6}$ for the $B(\psi(2S)\to D_s^-\pi^+)$, and $7.0\times10^{-6}$ for the $B(\psi(2S)\to D_s^-\rho^+)$ decays.

\section{Summary}
Based on $(2712.4\pm14.3)\times 10^6$ $\psi(2S)$ events collected at $\sqrt{s}=3.686\ \rm{GeV}$ with the BESIII detector~\cite{bes3:psipnum}, we search for the charmonium weak decays $\psi(2S)\to D_s^-\pi^+$ and $\psi(2S)\to D_s^-\rho^+$ for the first time. No significant signals are observed above the expected backgrounds. Upper limits of $B(\psi(2S)\to D_s^-\pi^+)$ and $B(\psi(2S)\to D_s^-\rho^+)$ are set to be $1.4\times 10^{-6}$ and $7.0\times 10^{-6}$ at the 90\% C.L., respectively. 
These limits are consistent with SM expectations, which predict the BFs at the level of $10^{-8}$ or lower.
The current upper limits on the branching fractions suggest that the present data sample lacks sufficient sensitivity to probe for potential new physics effects through these decay channels. Consequently, further data acquisition is highly motivated.

\section*{Acknowledgments}
The BESIII Collaboration thanks the staff of BEPCII (https://cstr.cn/31109.02.BEPC) and the IHEP computing center for their strong support. This work is supported in part by National Key R\&D Program of China under Contracts Nos. 2023YFA1606000, 2023YFA1606704; National Natural Science Foundation of China (NSFC) under Contracts Nos. 11635010, 11935015, 11935016, 11935018, 12025502, 12035009, 12035013, 12061131003, 12192260, 12192261, 12192262, 12192263, 12192264, 12192265, 12221005, 12225509, 12235017, 12342502, 12361141819; the Chinese Academy of Sciences (CAS) Large-Scale Scientific Facility Program; the Strategic Priority Research Program of Chinese Academy of Sciences under Contract No. XDA0480600; CAS under Contract No. YSBR-101; 100 Talents Program of CAS; The Institute of Nuclear and Particle Physics (INPAC) and Shanghai Key Laboratory for Particle Physics and Cosmology; ERC under Contract No. 758462; German Research Foundation DFG under Contract No. FOR5327; Istituto Nazionale di Fisica Nucleare, Italy; Knut and Alice Wallenberg Foundation under Contracts Nos. 2021.0174, 2021.0299, 2023.0315; Ministry of Development of Turkey under Contract No. DPT2006K-120470; National Research Foundation of Korea under Contract No. NRF-2022R1A2C1092335; National Science and Technology fund of Mongolia; Polish National Science Centre under Contract No. 2024/53/B/ST2/00975; STFC (United Kingdom); Swedish Research Council under Contract No. 2019.04595; U. S. Department of Energy under Contract No. DE-FG02-05ER41374

\newpage
M.~Ablikim$^{1}$\BESIIIorcid{0000-0002-3935-619X},
M.~N.~Achasov$^{4,b}$\BESIIIorcid{0000-0002-9400-8622},
P.~Adlarson$^{81}$\BESIIIorcid{0000-0001-6280-3851},
X.~C.~Ai$^{87}$\BESIIIorcid{0000-0003-3856-2415},
C.~S.~Akondi$^{31A,31B}$\BESIIIorcid{0000-0001-6303-5217},
R.~Aliberti$^{39}$\BESIIIorcid{0000-0003-3500-4012},
A.~Amoroso$^{80A,80C}$\BESIIIorcid{0000-0002-3095-8610},
Q.~An$^{77,64,\dagger}$,
Y.~H.~An$^{87}$\BESIIIorcid{0009-0008-3419-0849},
Y.~Bai$^{62}$\BESIIIorcid{0000-0001-6593-5665}
% ...
\\
\vspace{0.2cm}
(BESIII Collaboration)\\
\vspace{0.2cm} {\it
$^{1}$ Institute of High Energy Physics, Beijing 100049, People's Republic of China\\
% ...
$^{4}$ Budker Institute of Nuclear Physics SB RAS (BINP), Novosibirsk 630090, Russia\\
% ...
$^{31}$ INFN Sezione di Ferrara, (A)INFN Sezione di Ferrara, I-44122, Ferrara, Italy; (B)University of Ferrara, I-44122, Ferrara, Italy\\
% ...
$^{39}$ Johannes Gutenberg University of Mainz, Johann-Joachim-Becher-Weg 45, D-55099 Mainz, Germany\\
% ...
$^{62}$ Southeast University, Nanjing 211100, People's Republic of China\\
% ...
$^{64}$ State Key Laboratory of Particle Detection and Electronics, Beijing 100049, Hefei 230026, People's Republic of China\\
% ...
$^{77}$ University of Science and Technology of China, Hefei 230026, People's Republic of China\\
% ...
$^{80}$ University of Turin and INFN, (A)University of Turin, I-10125, Turin, Italy; (B)University of Eastern Piedmont, I-15121, Alessandria, Italy; (C)INFN, I-10125, Turin, Italy\\
$^{81}$ Uppsala University, Box 516, SE-75120 Uppsala, Sweden\\
% ...
$^{87}$ Zhengzhou University, Zhengzhou 450001, People's Republic of China\\

\vspace{0.2cm}
$^{\dagger}$ Deceased\\
% ...
$^{b}$ Also at the Novosibirsk State University, Novosibirsk, 630090, Russia\\
% ...

}

\end{document}